\documentstyle[aps,prb,floats,psfig]{revtex} 

\def\be{\begin{equation}}
\def\ee{\end{equation}}
\def\beq{\begin{eqnarray}}
\def\eeq{\end{eqnarray}}

\begin{document}
\twocolumn[\hsize\textwidth\columnwidth\hsize\csname@twocolumnfalse%
\endcsname

\draft

\title{Theory of the Temperature and Doping dependence of the Hall
  effect in a model with X-ray edge singularities in $ d = \infty $ }
\author{Mukul S. Laad and Stefan Blawid} 
\address{Max-Planck-Institut f\"ur Physik komplexer Systeme, 
         N\"othnitzer Stra\ss e 38, D-01187 Dresden, Germany}
\date\today 
\maketitle

\begin{abstract}
  We explain the anomalous features in the Hall data observed
  experimentally in the "normal" state of the high-$ T_{c} $
  superconductors.  We show that a consistent treatment of the local
  spin fluctuations in a model with x-ray edge singularities in $ d =
  \infty $ reproduces the temperature $ (T) $ and the doping
  dependence (x) of the Hall constant $ R_{\rm H} $ as well as the
  Hall angle in the normal state.  The model has also been invoked to
  justify the marginal Fermi liquid (MFL) behavior, and provides the
  first consistent explanation of the Hall anomalies for a non-Fermi
  liquid in $ d = \infty $.

\end{abstract}

]

\vspace*{0.1cm}
\section{INTRODUCTION}

The discovery of high-T$_c$ superconductors in Cu-O based compounds
has led to an upsurge of theoretical work concerning the unusual
normal state properties of these materials, which appear not to
conform to the framework of the Landau Fermi liquid theory [1, 2]. 

A way of unifying the diverse anomalies observed in experiment was
proposed by Varma and coworkers [2], who suggested a phenomenological
ansatz for the spectrum of charge and spin fluctuations. For low
frequencies $\omega \ll v_Fq$ this marginal-Fermi-liquid (MFL) ansatz
is \be Im \chi_{\rho, \sigma} (\omega) \sim \left \{ \matrix{ - \rho
  (0) \frac{\omega}{T} , & \omega \ll T , \cr - \rho (0) , & T \ll
  \omega \ll \omega_c} \right.  \ee where $\omega_c$ is a cut-off
energy. The s.p.~self-energy $\Sigma (\omega) \sim \omega \ln \omega
\pm i | \omega |$, as a consequence of (1), and this reconciles the
unusual normal state anomalies with the existence of the Luttinger
Fermi surface.

As emphasized in [2], the singularities in (1) are in the frequency
dependence; the momentum dependence is assumed smooth.  The exact
solution of the Falicov-Kimball model (FKM) in $d = \infty$, [3] where
local fluctuations are treated exactly, has been shown to lead to the
above spectrum near half-filling, and to a Fermi liquid phase at
farther fillings.  Furthermore, if these singularities do not depend
on any special symmetries which are lost in the lattice problem, they
are likely to survive in the lattice problem.

Varma et al.~[2, 4] have solved a multiband Hubbard model within the
impurity approximation to obtain the MFL form for the local
susceptibilities. However, the MFL theory has not been able to account
for the $T$ and doping ($x$) dependence of the Hall constant $R_{\rm
  H}$.  Thus, the reconciliation of the Hall anomalies within the MFL
hypothesis is clearly an open issue of current interest.

The tomographic Luttinger liquid-based model of Anderson and Ong
provides a consistent explanation of the anomalous features seen in
the d.c as well as a.c Hall effect [5].  It is, however, based on an
extension of 1d ideas to the 2d case; such an extension is a hotly
debated issue, and there is no rigorous proof yet, inspite of intense
efforts [6].

The calculation of the Hall constant for strongly correlated metals is
especially hard to describe.  It involves the computation of the
conductivity tensor to first order in the magnetic field.  This
involves the rather hard problem of computing a three-point function.
In addition, the calculation of the vertex corrections for strongly
interacting lattice fermion models is a formidable technical problem
that has not been attempted.  The above reasons make it imperative to
develop techniques where some of the difficulties maybe circumvented
without sacrificing essential correlation effects.
 
The unusual $ T $ and $ x $ dependence of the Hall effect in the
high-$ T_{c} $s has been recognised to be a striking anomaly of the
normal state, and cannot be understood within conventional Fermi
liquid ideas [7-10].  In conventional metals, it is rare [8] to
observe a T-dependence of $ R_{\rm H} $ above a fraction (0.2 to 0.4)
of the Debye temperature $ \Theta_{\rm D} $.  However, in the
hole-doped cuprates, the Hall coefficient (with $ {\bf B} $ along $
{\bf c}) $ shows a decrease with increasing T upto room temperature,
going roughly like $ 1/(T+C) $, where $ C $ is a constant.
Suppression of superconductivity by doping with impurities or by
moving away from optimal doping suppresses the T-dependence of $
R_{\rm H} $.  There is no effect of phonons, in contrast to the
situation in normal metals.  The Hall angle $ \mbox{cot}\theta_{\rm H}
$ goes like $ \alpha T^{2}+\beta(x) $ [8].  The persistence of these
anomalies to a rather high temperature precludes scattering mechanisms
involving phonons [10] or due to anisotropic Fermi surfaces [11].

In this paper, we examine the behavior of the Hall coefficient $
R_{\rm H} $ as well as the Hall angle $\mbox{cot}\theta_{\rm H}$ by an
exact treatment of the local spin fluctuations in the uniform phase of
the FKM, which exhibits x-ray edge (XRE) singularities in $ d = \infty
$.  Our calculation presents a consistent explanation of the
experimentally observed Hall anomalies, and is, we believe, the first
microscopic calculation of the Hall anomalies for a non-FL metal in $
d = \infty $.

\section{THEORETICAL FRAMEWORK: MODEL AND LOCAL SPECTRAL DENSITY}

In what follows, we consider the Falicov-Kimball model as an effective
model describing the electronic degrees of freedom in the ${\rm
  CuO}_{2}$ layers in the cuprate SCs.  The Falicov-Kimball model in
2d in the large $U$ limit is defined by, \be H = - t \sum_{<ij>}
(c^\dagger_{i} c_{j} + h.c) + U \sum_i n_{ic} n_{id} - \mu
\sum_{i}(n_{ic} + n_{id}) \ee In this model, $t$ and $U$ should be
understood as effective parameters which are determined by comparison
of the low-energy spectra of (2) with that of a more realistic
three-band model [2].  The proposed FKM bears some similarity to the
effective model of ref.~[2].  In ref.[2], the authors started with a
full three band Hubbard model involving the complete local structure
of a unit Cu-O cell embedded within the 2d square Cu-O plane.  In the
strong correlation limit the low lying eigenstates were retained, and
the resulting hamiltonian was transformed to essentially the FKM (with
$ U = \infty $ in our case).  Another justification comes from the
proposal of Anderson [1], who suggests that the non-FL anomalies in
the normal state of cuprate SCs arises from effects akin to the
orthogonality catastrophe in the x-ray edge problem.  As remarked
above, it is unclear whether the 2d Hubbard model can exhibit these
phenomena.  It is, nevertheless important to separate the effects
coming from the x-ray edge-like physics from those arising due to low
dimensionality.  This enables us to separate out the effects of
bandstructure (like van-Hove singularities) and investigate the extent
to which strong local correlations (which are treated exactly in $ d =
\infty $) are responsible for the anomalous normal state behaviors.
 
We are able to address the first issue in this paper, since the FKM
explicitly shows the x-ray edge (XRE) behavior in $ d = \infty $.  We
mention that such calculations for the Hubbard model in $ d = \infty $
have been carried out by two groups [12, 13], both of whom obtain a
crossover to a Fermi liquid (FL) below a certain crossover scale $
T_{coh} $.  This is not surprising in view of the fact that the
one-band Hubbard model in $ d = \infty $ always has a FL paramagnetic
metallic phase [14].  In contrast, the metallic state of the FKM near
half-filling is a non-FL, with a crossover to a FL at higher dopings
[15].

We have solved the FKM exactly in the $ d = \infty $ limit by an
equation-of- motion technique [15]; this reproduces exactly the known
exact solution of this model in this limit [15].  In what follows, we
compute the conductivity tensor exactly in $ d = \infty $; this is
because of the remarkable fact that the vertex corrections to the
conductivity vanish identically [16] in $ d = \infty $.  This
simplifying feature is what makes the calculation of the
conductivities and of the Hall effect exact in this limit.

We have solved the lattice model exactly in $ d = \infty $.  Since the
authors of ref.[2] solve an impurity model, they require fine-tuning
of parameters to reach the critical point.  Since we perform a lattice
calculation exact in $ d = \infty $, the critical behavior survives
for a finite range of filling, as the authors of ref.~[2] anticipate.
In this FKM [2,15], the $ d $ holes are immobile. This means that
$[n_{id}, H] = 0 \forall i$; hence, the model has an exact local
$U(1)$ symmetry.  We notice that the model eqn (2) is different from
the usual Hubbard model, which has a global U(1) symmetry associated
with ${\it total}$ fermion number conservation.  We show below that
the system develops extra singularities in a magnetic field, in
addition to those implied by the XRE physics, of a form necessary to
explain the anomalous Hall data.

The exact computation of the local c-fermion single particle
propagator in $ d = \infty $ follows the steps in ref.~[10], and
yields (for a Lorentzian unperturbed DOS $\rho_o (z) = (\Delta / \pi)
(z^2 + \Delta^2)^{-1}$, $ \Delta $ being the effective bandwidth in
the unperturbed problem) \be G^{c}_{ii} (i\omega_\ell) =
\frac{1}{2\pi} \left[\frac{1-n_{d}}{i\omega_\ell + i \Delta \mbox{sgn}
  \omega_\ell} + \frac{n_{d}}{i\omega_\ell - U + i\Delta \mbox{sgn}
  \omega_\ell}\right] \ee with the self-energy \be \Sigma_{c}
(i\omega_\ell) = Un_{d} + \frac{U^2 n_{d} (1 - n_{d})}{i\omega_\ell -
  U (1 - n_{d}) + i \Delta \mbox{sgn} \omega_\ell}\;.\ee Also, it is
easily seen that \be \langle\langle n_{id} c_{i}; c^\dagger_{i}
\rangle\rangle_{i\omega_\ell} = \frac{n_{d}}{2\pi}
\frac{1}{i\omega_\ell - U + i\Delta \mbox{sgn} \omega_\ell} \;.\ee The
s.p.~and the two-particle local spectral densities are \be \rho_{c}
(\omega) = \frac{\Delta}{2\pi^2} \left[\frac{1-n_{d}}{\omega^2 +
  \Delta^2} + \frac{n_{d}}{(\omega - U)^2 + \Delta^2}\right] \ee and
\be \rho^{(2)} (\omega) = \frac{\Delta}{2\pi^2} \frac{n_{d}} {(\omega
  - U)^2 + \Delta^2} \;.\ee From eqns (6) and (7), it is clear that
the low-energy spectrum is a superposition of s.p.~and two-particle
states, leading to a breakdown of Landau's Fermi liquid theory.

\section{COMPUTATION OF THE HALL EFFECT}

The computation of the Hall constant involves evaluation of the
conductivity tensor $ \sigma_{\alpha\beta}(\omega = 0) $.  In the $ d
= \infty $ approximation the transport coefficients do not involve the
vertex corrections [16] and the dc conductivity is given by \be
\sigma_{xx}(0) = c_{xx}\int d\epsilon\rho_{0}(\epsilon)\int d\omega
A^{2}(\epsilon,\omega)\beta\frac{\mbox{sech}^{2}(\beta\omega/2)}{4}
\ee where $ c_{xx} = e^{2}\pi/(d\hbar a_{0}) $, $ a_{0} $ is the
lattice spacing, $ \rho_{0}(\epsilon) $ is the bare (lorentzian)
density of states (DOS) and $ A(\epsilon,\omega) =
-(1/\pi)Im[\omega+\mu-\epsilon-\Sigma(\omega)]^{-1} $ is the s.p.~
spectral function.  The Hall conductivity is given by [17] \be
\sigma_{xy}(0) = Bc_{xy}\int d\epsilon\rho_{0}(\epsilon)\epsilon\int
d\omega
A^{3}(\epsilon,\omega)\beta\frac{\mbox{sech}^{2}(\beta\omega/2)}{4}
\ee where $ c_{xy} = |e|^{3}\pi^{2}a_{0}/3d^{2}\hbar^{2} $.  The Hall
constant and the Hall angle are then given by $ R_{\rm H} =
\sigma_{xy}(0)/B\sigma_{xx}^{2}(0) $ and $ \mbox{cot}\theta_{\rm H} =
\sigma_{xx}(0)/B\sigma_{xy}(0)$.  We see that the Hall constant enters
to zeroth order in $ 1/d $ inspite of the conductivities entering to
order $ 1/d $ and $ 1/d^{2} $ respectively (see eqns above).
  
In the remainder, we use the spectral density eqn (6) to compute the
Hall constant and the Hall angle as a function of the doping
concentration $ x $ and temperature $ T $.  We work with $ U/\Delta =
4.0 $, a value representative of the strongly correlated case that
also allows us to compare our results with those obtained for the
Hubbard model with the same parameters by Majumdar et al.~ [12].  We
choose values of the hole concentration to illustrate the
qualitatively different regimes; (1) $ x = 0.1 $, $ x = 0.2 $, to
describe the $ {\it optimally} $ doped regime, and (2) $ x = 0.32 $,
the $ {\it overdoped} $ regime.  We will further work in the quantum
paramagnetic case ($n_{c} = n_{d}$ [18]). Thus, our results should
apply best to these cases where there is no remnant of long-range
order. To treat the underdoped case, one has to consider short-range
AFM fluctuations, which grow as one approaches the limit of
half-filling.  This would require us to consider possible ground
states with broken symmetries, and is beyond the scope of this work.
In addition to antiferromagnetic fluctuations, the effects of $ {\it
  disorder} $ are also dominant in the underdoped regime, and this
requires a reanalysis with the inclusion of disorder effects in a MFL
[19].

The local spectral density of the model exhibits a two subband
structure with the characteristic transfer of spectral weight of
excitations from the $ {\it upper} $ to the $ {\it lower} $ Hubbard
band upon hole doping. The inset to FIG.\ref{abb.dot} shows the
evolution of the chemical potential $ \mu $ determined from the eqn $
n = 1-x =n_{d}+\int_{-\infty}^{\mu}\rho_{c}(\omega)d\omega $ with hole
doping.  It is important, for what follows, to emphasize the fact that
$ \mu $ is in the region of the overlapping s.p.~and two-particle
spectral densities (see eqns (6)-(7)) for case (1), while it has
clearly moved out of this region into the s.p.~part of the spectral
density for the overdoped case (2).

\section{RESULTS AND DISCUSSION}

\begin{figure}
\centerline{\psfig{figure=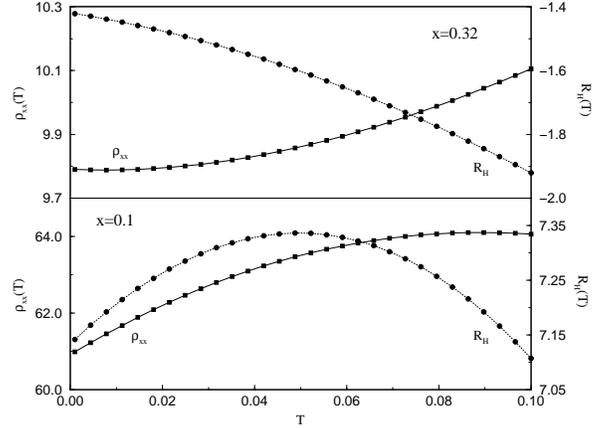,width=0.46\textwidth,angle=-90}}
\caption{\label{abb.join}
  Resistivity $ \rho_{xx}(T) $ and $ R_{\rm H}(T) $ as a function of T for
  two values of doping ($ x_{1} = 0.32 $ (upper) \& $ x_{2} = 0.1 $
  (lower Panel)).  Notice the change of slope in the resistivity from
  sublinear to quadratic as also a sign change in $ R_{\rm H}(T) $ with
  increasing $ x $.  }
\end{figure}
\begin{figure}
\centerline{\psfig{figure=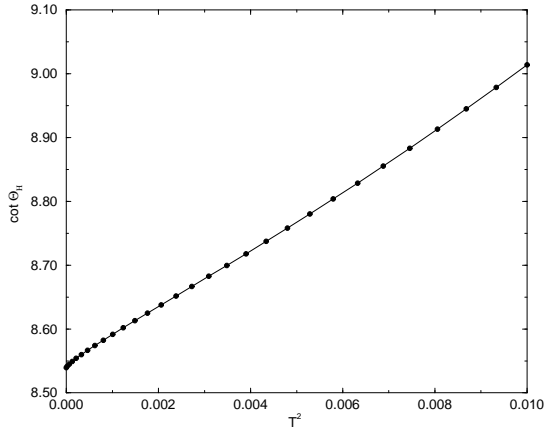,width=0.46\textwidth,angle=-90}}
\caption{\label{abb.hallw}
  Hall angle $\mbox{cot}\theta_{\rm H}(T) $ at $ x = 0.1 $.  It goes as $ g(T)
  = aT^{2}+b $ over a wide temperature range.}
\end{figure}
\begin{figure}
\vspace*{-1.5cm}
\centerline{\psfig{figure=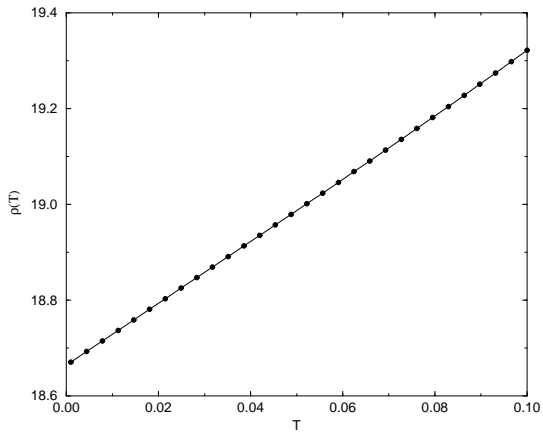,width=0.46\textwidth,angle=-90}}
\caption{\label{abb.res} 
  Resistivity $ \rho_{xx}(T) $ for $ x = 0.2 $.  The clean linear
  behavior over a very wide temperature scale is clear.  }
\end{figure}
\begin{figure}
\vspace*{-1.5cm}
  \centerline{\psfig{figure=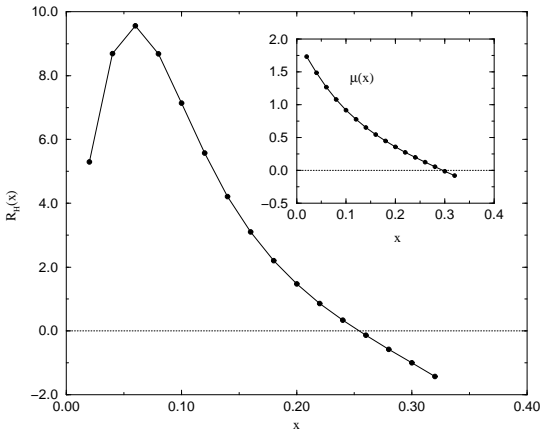,width=0.46\textwidth,angle=-90}}
  \caption{\label{abb.dot}
    Hall constant $ R_{\rm H}(x) $ as a function of $ x $.  It changes
    sign at around the $ x_{c} $ corresponding to the MFL-FL
    crossover with doping.  }
\end{figure}

In FIG.\ref{abb.join}, we show the field-free inplane resistivity $
\rho_{xx}(T) $ as well as the Hall constant $ R_{\rm H} $ for the
cases (1) \& (2) above.  We also show the $ T $-dependence of the Hall
angle for the case (1) in FIG.\ref{abb.hallw}.  In FIG.\ref{abb.res}
we show the striking linear-in-$ T $ dependence of $ \rho_{xx}(T) $
for $x = 0.2$ over a wide temperature scale.  This linear behavior,
observed near optimal doping, has been regarded as one of the signals
of the non-Fermi liquid metallic normal state of the cuprate SCs.
Remarkably, we see that the essential features of the $ T $-dependence
of $ R_{\rm H} $ as well as $\mbox{cot}\theta_{\rm H} $ are reproduced
in qualitative agreement with experimental results [7-10].  More
remarkably, in the FIG.\ref{abb.dot}, we see that $ R_{\rm H} $ as a
function of doping changes sign and becomes (for hole doping) negative
at around the $ x_{c} $ corresponding to the MFL to FL crossover.  The
inplane resistivity also goes quadratically with $ T $ for $ x = 0.32
> x_{c} $, (FIG.\ref{abb.join}) revealing the doping-induced crossover
to a FL. Note however, that our approach does not allow us to study
the evolution of these quantities close to half-filling as remarked
above.

We reproduce qualitatively the correct $ T $-dependence for $ R_{\rm
  H} $ as well as $\mbox{cot}\theta_{\rm H} $ observed in the
optimally doped case.  $ R_{\rm H} $ goes roughly like $ 1/T $ at
intermediate $ T $, followed by a tendency to saturate as $ T_{c} $ is
approached from above.  The Hall angle $\mbox{cot}\theta_{\rm H} $
follows a $ T^{2} $ dependence over a wide temperature scale.  This
means that the Hall conductivity $ \sigma_{\rm H} $ goes like $
1/T^{3} $, again consistent with observations [8].  The
magnetoconductance $ \Delta\sigma_{xx} \simeq
\sigma_{xx}(\omega_{c}\tau_{\rm H})^{2} $, and so goes like $ 1/T^{5}
$ (see the ref.[20] and references therein).  The fact that the $
1/\tau_{\rm H} $ goes like $ T^{2} $ leads also to a microscopic
justification for the idea of two relaxation times in the normal state
of cuprate SCs.  The connection to Anderson's proposal of two
relaxation rates in the cuprate materials is suggestive, and our
calculation shows that the anomalies are linked to the XRE-like
physics inherent in the FKM in $ d = \infty $.  The XRE model in any
dimension can be bosonized [1] to a Tomonaga-Luttinger-like model on a
half-line, leading to asymptotic spin-charge separation.  Thus, our
results represent a higher-dimensional realization of Anderson's
ideas.
 
Earlier studies have developed this idea within a phenomenological
framework using the Boltzmann equation [20].  The skew-scattering
phenomenology requires near perfect particle-hole symmetry [21], while
the other approach [20] invokes two scattering processes that are
either even or odd under the charge-conjugation operator.  Anderson
[5] has provided a consistent explanation of the Hall anomalies within
the framework of the tomographic Luttinger liquid hypothesis.
However, it has not so far been shown conclusively [6,22,23] that the
Luttinger liquid concept can be extended to two dimensions.  Recently,
Mahesh et al.~[24] have computed the $T$ and $x$ dependence of $R_{\rm
  H}$ by numerical diagonalization of the one-band Hubbard model on
finite-sized clusters. They were able to account for the anomalous
behavior of $R_{\rm H}$.  We have provided an alternative explanation
for the magnetotransport anomalies within the $ d = \infty $
approximation, which for the effective Falicov-Kimball model also
leads to a non-Fermi liquid near $ n = 1 $.

It is interesting to compare our results with those obtained for the
Hubbard model by Majumdar et al.~[12], who studied the Hubbard model
with a weak next-nearest-neighbor (n.n.n) hopping strength $ t' $ in $
d = \infty $ using the iterated perturbation theory away from
half-filling.  They studied the $ T $-dependence of the usual Hall
constant, as well as that of the infinite-frequency Hall constant $
R_{\rm H}^{*} $ [25].  This latter quantity does not depend on the
low-energy structure in the s.p.~spectrum, but measures only the
effect of high-energy processes.  Majumdar et al.~found, interestingly
enough, that the $ T $-dependence of $ R_{\rm H}^{*} $, rather than
that of $ R_{\rm H} $, mimics the experimentally observed behavior.
While this result can be taken as evidence, along with indications
from other probes [26], of the importance of local fluctuations, it is
known that the Hubbard model, with or without n.n.n hopping, always
yields a $ T = 0 $ paramagnetic Fermi liquid metallic phase in $ d =
\infty $ [27].  A natural explanation of the observation made in [12]
arises within our calculation.  In contrast to the situation in the
Hubbard model, the spectrum at low energies in the FKM is
scale-invariant, there is no coherent FL-like feature at low-$ T $,
and the high-energy incoherent features in the spectral density are
pulled down to low energy.  Close to half-filling, the chemical
potential $ \mu $ is pinned in the region where the spectrum is a
superposition of the low-energy s.p.~and the high-energy two-particle
states, and so the transport is dominated by anomalous high-energy
scattering processes.  With increasing deviation from the $ n = 1 $
case, $ \mu $ shifts out of this region (inset of the
FIG.\ref{abb.dot}) into the s.p.-dominated part of the spectrum,
leading to the emergence of a more conventional behavior, as evidenced
in the change of sign in $ R_{\rm H}(x) $ as a function of doping
(FIG.\ref{abb.dot}).

A problem with the calculations presented here is the finite $ T = 0 $
intercept in $ \rho_{xx}(T) $ as well as $\mbox{ cot}\theta_{\rm H}(T)
$.  The resistivity $ \rho_{xx}(T) = \rho_{0} + \alpha T $ where $
\alpha $ decreases with increasing $ \Delta $ or $ x $, while
$\mbox{cot} \theta_{\rm H}(T) = aT^{2}+b $.  This is indicative of
residual elastic scattering processes, resulting from the quenched d
"impurities" at $ T = 0 $ in our calculations.  This is the artifact
of the $ d = \infty $ limit, and finite dimensional extensions are
required to remedy this situation.  Physically, the effects of static
disorder could lead to such effects, as seen in experiments.  We have
not done this here, however.

\section{CONCLUSION}

To conclude, we have presented the first explicit calculation of the
Hall anomalies observed experimentally in the normal state of cuprate
SCs.  We have shown that the observed Hall anomalies as a function of
temperature and doping, along with those seen in magnetotransport, can
be understood simply in terms of the scale-invariant, non-FL spectrum
characteristic of a model exhibiting XRE-like singularities in $ d =
\infty $.  It has also been possible to show [3] that the exact
solution of the FKM in $ d = \infty $ has a non Fermi liquid (NFL)
metallic phase near $ n = 1 $, and so our results also represent a
first microscopic calculation of the Hall anomalies for a strongly
correlated non-FL metal.\\ 
  
\section{ACKNOWLEDGEMENTS} 

One of us (Mukul S.~Laad) wishes to thank Prof. P.~Coleman for a
discussion on non-FL behavior in the XRE model.  M.~S.~Laad also
acknowledges financial support from the Alexander von-Humboldt
Foundation and thanks Prof.~P.~Fulde for advice and hospitality at the
MPI, Dresden.\\ 
 
\noindent{\large{References}}\\
$[1]$ P.W. Anderson and Y. Ren, in ``Superconductivity - The Los
Alamos Symposium'', eds. K. Bedell, et al., Addison Wesley.\\ $[2]$
C.M. Varma, P. Littlewood, S. Schmitt-Rink, E. Abrahams and A.E.
Ruckenstein, {\it Phys. Rev. Lett.} {\bf 63}, 1996 (1989); also C.
Sire, C.M. Varma, A.E. Ruckenstein and T. Giamarchi, {\it Phys. Rev.
  Lett.} {\bf 72}, 15 (1994).\\ $[3]$ Mukul S.Laad, accepted for
publication in Mod. Phys.  Lett. B\\ $[4]$ C. Sire, C. M. Varma, A.
E.Ruckenstein and T. Giamarchi, {\it Phys. Rev. Lett.} {\bf 72}, 15
(1994).\\ $[5]$ P. W. Anderson, Physical Review Letters {\bf 67}, 2092
(1991).\\ $[6]$ J. R. Engelbrecht and M. Randeria, {\it Phys. Rev.
  Lett.} {\bf 65}, 1032 (1990).\\ $[7]$ See for e.g, N. P. Ong, in
{\it Physical Properties of the High Temperature Superconductors},
edited by D. M. Ginsberg (World Scientific, Singapore 1990), vol.2, p.
459.\\ $[8]$ P. Chaudhuri et al., Phys. Rev. {\bf B 36}, 8903
(1987).\\ $[9]$ T. R. Chien, D. A. Brawner, Z. Z. Wang, and N. P. Ong,
Phys. Rev. {\bf B 43}, 6242 (1991).\\ $[10]$ See for e.g, N. P. Ong,
T. W. Jing, T. R. Chien, Z. Z.  Wang, T. V. Ramakrishnan, J. M.
Tarascon and K. Remschnig, Physica {\bf C 185-189}, 34 (1991).\\ 
$[11]$ J. M. Ziman, {\it Electrons and Phonons}, Oxford University
Press (1960).\\ $[12]$ P. Majumdar and H. R. Krishnamurthy, cond-mat
{\it 9512151}.\\ $[13]$ T. Pruschke, {\it et.al} Adv. Phys. {\bf 44},
187 (1995).\\ $[14]$ See A. Georges, G. Kotliar, W. Krauth and M.
Rozenberg, {\it Revs. Mod.  Phys.}, {\bf 68}, 13 (1996).\\ $[15]$ Q.
Si, A. Georges and G. Kotliar, {\it Phys. Rev.} {\bf B 46}, 1261
(1992).  see also Mukul S.Laad, {\it Phys. Rev.} {\bf B 49}, 2327
(1994).\\ $[16]$ A. Khurana, {\it Phys. Rev. Lett.}, {\bf 64}, 1990
(1990).\\ $[17]$ P. Voruganti, A. Golubentsev and S. John, {\it Phys.
  Rev.} {\bf B 45}, 13945 (1992).\\ $[18]$ H. Fukuyama and H.
Ehrenreich, Phys. Rev. B${\bf 7}$, 3266 (1973), see also F. Brouers
and F. Ducastelle, J. de Phys. ${\bf 36}$, 851 (1975).  The Hubbard
III approximation used by these authors becomes exact for the FKM in
$d=\infty$ and yields $m=0$ ($n_{c}=n_{d}$) in our case.\\ $[19]$ G
.Kotliar, E. Abrahams, A. E. Ruckenstein, C. M. Varma, P. B.
Littlewood and S. Schmitt-Rink, {\it Europhysics Letters}, {\bf 15}
(6), 655 (1991).\\ $[20]$ G. Kotliar, A. Sengupta and C. M. Varma,
{\it Phys.  Rev.} {\bf B 53}, 3573 (1996).\\ $[21]$ P. Coleman,{\it
  Phys. Rev. Lett.} {\bf 76}, 1324 (1996).\\ $[22]$ P.W. Anderson,
{\it Phys. Rev. Lett.} {\bf 64}, 1839 (1990).\\ $[23]$ See for e.g, P.
W. Anderson, in {\it The Hubbard Model}, Edited by D. Baeriswyl et
al., Plenum Press, New York (1995).  See also J. R. Engelbrecht and M.
Randeria, {\it Phys. Rev. Lett.} {\bf 65}, 1032 (1990).\\ $[24]$ P.
Mahesh and B.S. Shastry, unpublished\\ $[25]$ B. S. Shastry, B. I.
Shraiman and R. R. P. Singh, {\it Phys. Rev. Lett.} {\bf 70}, 2004
(1993).\\ $[26]$ C. M. Varma, {\it Int. J. Mod. Phys.} {\bf B3}, 12,
2083 (1989).\\ $[27]$ See A. Georges, G. Kotliar, W. Krauth and M.
Rozenberg, {\it Revs. Mod. Phys.}, {\bf 68}, 13 (1996).\\ 

\end{document}